# Multilayered Maxwell's fisheye lens as waveguide crossing

M. M. Gilarlue[a], J. Nourinia[a], Ch. Ghobadi[a], S. Hadi Badri[b,*], H. Rasooli Saghai[c]

[a] Department of Electrical Engineering, Urmia University, Urmia 57153, Iran

[b] Department of Electrical Engineering, Azarshahr Branch, Islamic Azad University, Azarshahr, Iran

[c] Department of Electrical Engineering, Tabriz Branch, Islamic Azad University, Tabriz, Iran

Corresponding author.

* E-mail addresses: sh.badri@iaut.ac.ir

## Abstract

The Maxwell's fisheye (MFE) lens, due to its focusing properties, is an interesting candidate for implementing the crossing of multiple waveguides. The MFE lens is implemented by two different structures: concentric cylindrical multilayer and radially diverging gourd-shaped rods. Realization of the refractive index profile of the lens is achieved by controlling the thickness ratio of the alternating Si and $SiO_2$ layers determined by effective medium theory. Both structures are optimized to cover the entire C-band in the single mode implementation. The transmission efficiency of the ring-based structure is superior to the radial-based implementation, however, the radial-based structure almost covers the entire U-band as well. Other communication bands are partially covered in both cases. Full-wave simulations prove that the performance of multimode waveguide crossing based on the MFE lens with a radius of $2.32 \mu m$ is promising with the average insertion loss of *0.17dB* and crosstalk levels below *-24.2dB* in the C-band for $TM_0$ and $TM_1$ modes. The multimode intersection almost covers the entire C, L, and U bands of optical communication.

## 1. Introduction

The crossing of multiple waveguides at one intersection is inevitable in the highly dense photonic integrated circuits (PICs) where numerous devices are integrated in a very small area. An intersection with desired performance should have low insertion loss, low reflection, low crosstalk, and broad bandwidth. Various methods have been proposed for designing the intersection of two waveguides for photonic crystals (PhCs) including utilization of single defect with doubly degenerate modes in square lattice [1]. The intersection designed with this method has narrow bandwidth and simultaneous crossing of two signals with the same wavelength causes interference [2]. To overcome these limitations coupled cavity waveguides were proposed in triangular lattice [2]. Utilization of coupled cavity waveguides generates leakage in vertical direction i.e. out of the plane of photonic crystal slab [3]. Other methods include topology optimization [4], Wannier basis design [5], and bridged waveguide intersection [6]. The design of PhC waveguide crossings is generally based on cavity resonators which have inherently narrow bandwidth. The bandwidth can be increased by decreasing the Q-factor of the cavity resulting in lower transmission of the crossing. Nevertheless, the cross-talk levels of these designs are typically lower than -30*dB*. For silicon-on-insulator (SOI) waveguide crossings, different methods based on multimode interference [7, 8], mode expander [9], and wavefront matching [10] have been reported. Wavefront matching method has a very large footprint compared with multimode interference methods. The waveguide crossing mechanism based on impedance matched metamaterial have also been proposed [11]. This method has low bandwidth since it is based on negative refractive index metamaterial. To the best our knowledge, crossing of multiple waveguides at single intersection and multimode waveguide crossing have not been reported with methods of [1-6, 9-11]. New methods supporting the crossing of multiple waveguides based on inverse-design [12] and the MFE lens [13] have been investigated. Multimode waveguides play an important role in PICs and variety of devices are proposed based on them such as optical filters [14-16], power splitters [17, 18], polarization splitters [19, 20], sensors [21, 22], (de)multiplexers [23-26], switches [27-29], and wavelength converter [30]. Therefore, designing multimode waveguide crossings is vital in PICs and they have been introduced based on multimode-interference couplers [31, 32], subwavelength asymmetric Y-junction [33], and the MFE lens [34]. There has been no report of multiple waveguides crossing at an intersection with methods of [31-33]. However, the MFE lens supports the crossing of multiple multimode waveguides at an intersection with high bandwidth.

In this paper, it is demonstrated that the intersection of three single-mode or multimode waveguides can be implemented with MFE lens for transverse magnetic (TM) mode. Conventionally in PhCs, the TM mode is characterized by the electric field intensity being parallel to the rods and perpendicular to the plane of electromagnetic wave propagation and the transverse electric (TE) mode is characterized by the electric field intensity being parallel to the plane of electromagnetic wave propagation [35]. This paper presents, for the first time to the authors' knowledge, the implementation of MFE lens with concentric ring-based and radial-based structures instead of graded photonic crystals [13, 34]. The $Si/SiO_2$ multilayer with high refractive index contrast is adopted to implement the refractive index profile of the MFE lens. It is possible to realize metamaterials with intermediate refractive indices of the two constituent materials with the multilayer structure. The $Si/SiO_2$ is compatible with Complementary-Metal-Oxide-Semiconductor (CMOS) technology and reduces the cost of PIC development [36].

The refractive index of the MFE lens is given by [34]

$$n_{lens}(r) = \frac{2 \times n_{min}}{1+(r/R_{lens})^2} \quad , \quad (0 \leq r \leq R_{lens}) \tag{1}$$

where $n_{min}$ is the minimum refractive index of the lens at its edge, $R_{lens}$ is the radius of the lens and $r$ is the radial distance from the center of the lens. In order to minimize the reflection at the lens interface, the $n_{min}$ should be equal to the environment refractive index [37]. Radiation of the point source on the surface of the lens, due to the lens's spatial variation of the refractive index, is focused on the diametrically opposite point of the lens. Fig. 1(a) shows the simple intersection of three waveguides. The crosstalk is substantial in this case and the signal is very weak in the intended output waveguide. In Fig. 1(b) the concept of intersection based on the ring-based multilayer MFE is shown. In gradient index (GRIN) lenses light propagates along a curved trajectory. In these figures, power streamlines are used to visualize the average power flow.

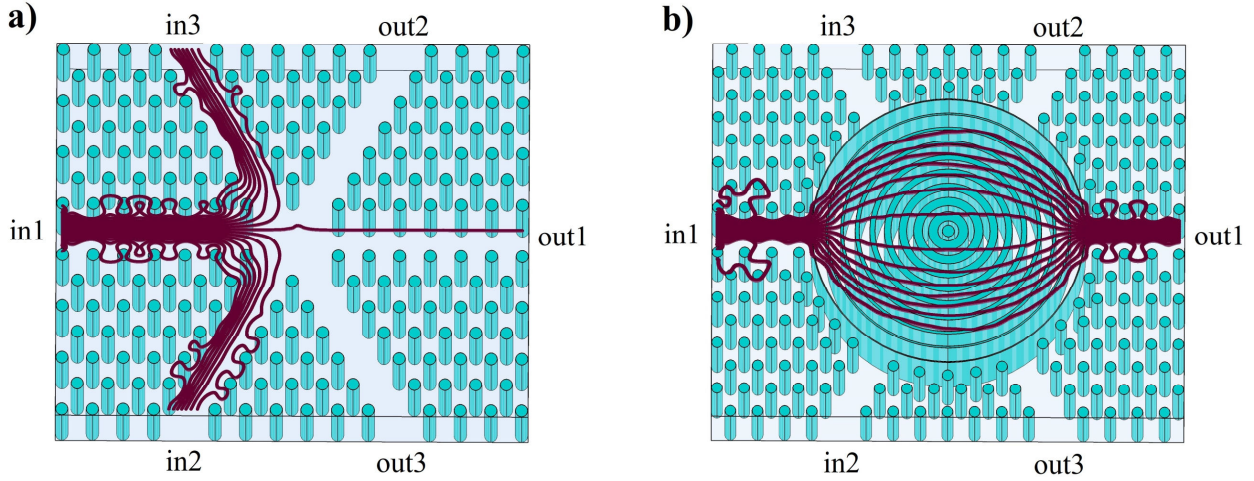

Fig. 1. Design concept. The flow of power is illustrated in the simple intersection of three waveguides in (a) and similarly with multilayer MFE lens as crossing medium in (b). The rods and inclusion layers of Si are embedded in SiO$_2$ background.

## 2. Multilayer GRIN lens design

The effective medium theory (EMT), also known as the effective medium approximation (EMA), is a homogenization method for determining the optical properties of composite materials based on the volume averaged fields. In the long-wavelength limit, where the wavelength of operation is considerably larger than the structure size, the wave does not resolve the subwavelength structure. In addition, Mie scattering and Bragg diffraction are negligible when the width of the layers and the duty cycle of the layered structure is well below the wavelength of the incident wave [38]. When these conditions are met, the structure can be considered as a homogeneous medium. The effective refractive index of the two-phase composite material depends on the ordered arrangement of the inclusions in the host to the electric field direction [39-42]. When the inclusions are arranged parallel to the electric field, the effective refractive index is approximated by [39]

$$n_{eff,TM}^2 = f_{inc}n_{inc}^2 + (1-f_{inc})n_{host}^2 \qquad (2)$$

where $n_{inc}$, $n_{host}$, and $n_{eff,TM}$ are the refractive indices of the inclusion, host, and effective medium for TM mode, respectively. And filling factor, $f_{inc}$, is the fraction of the total volume occupied by inclusions. When the inclusions are arranged perpendicular to the electric field, the effective refractive index is given by [39]

$$n_{eff,TE}^2 = \frac{n_{inc}^2 n_{host}^2}{f_{inc} n_{host}^2 + (1-f_{inc}) n_{inc}^2} \tag{3}$$

where $n_{eff,TE}$ is the refractive index of the effective medium for TE mode. When the comprising materials are isotropic, the dependence of the effective refractive index on the polarization is called form birefringence [43] and has interesting applications [44-46]. Form birefringent structures have high refractive index difference of TE and TM polarizations, $\Delta n$, compared to the naturally birefringent materials. Furthermore, the magnitude of $\Delta n$ can be adjusted by the duty cycle and the shape of the structure [47].

Multilayer structures could be used to implement GRIN medium [36], metamaterials [48, 49], hyperlenses [50, 51], or invisibility cloaks [52, 53]. It is worth mentioning that there is no natural material exhibiting cylindrical anisotropy, however, it can be realized with multilayer metamaterials composed of isotropic materials [51]. The EMT can be used to calculate the refractive index of the multilayer structures ensuring that the thickness of each layer is much less than the incident wavelength and a sufficient number of layers is used [52]. By varying the filling factor of the inclusion material it is possible to control the effective refractive index of the structure. In this paper, the MFE lens is implemented by two different multilayer structures with homogeneous isotropic materials. The first structure is the ring-based multilayer MFE lens composed of concentric rings alternatively being Si or $SiO_2$ with a gradually decreasing filling factor. Here Si and $SiO_2$ are considered as the inclusion and host, respectively. Using the same materials for implementation of the PhC and MFE lens greatly simplify the fabrication process [54]. Our procedure for designing the ring-based MFE lens is to divide it into concentric cylindrical layers of equal width. For calculating the width of inclusion annulus in the i-th layer Eq. 2 is rearranged as

$$f_{inc} = \frac{n_{eff,TM}^2 - n_{host}^2}{n_{inc}^2 - n_{host}^2} \tag{4}$$

The filling factor for the i-th layer is $f_{inc} = A_{inc}/A_i$ where $A_{inc} = 2\pi r_i dr_{inc}$ and $A_i = 2\pi r_i dr$ are areas of inclusion and the i-th layer, respectively. The width of the layers, $dr$, is kept constant in our design. Consequently, the width of the inclusion layer, $dr_{inc}$, is calculated by the following equation:

$$dr_{inc} = \frac{n_{eff,TM}^2 - n_{host}^2}{n_{inc}^2 - n_{host}^2} dr \tag{5}$$

For the first layer $f_{inc} = \pi dr_{inc}^2 / \pi dr^2$ is used instead of Eq. 5. This procedure can also be applied to the design of lamellar multilayer GRIN medium.

Similar to the previous method, in the design of radial-based structure, the lens is divided into concentric cylindrical layers of equal width. Then each layer is divided into the desired number of equal annular sectors. In each annular sector, the filling factor is realized by inclusion of a sector with the same radii but smaller central angle. The inclusion sector is located at the center of the annular sector. Then the corners of consecutive inclusion sectors lying on the same radius are interpolated to form a radial rod.

## 3. Numerical simulation and discussion

The two-dimensional (2D) simulations were performed with Comsol Multiphysics™ to verify the MFE lens's performance. The intersection of channel waveguides based on MFE lens is investigated in [34]. In our design, PhC waveguides are considered, therefore, the MFE lens is surrounded by PhC structure. Consequently, optical waves can only pass through the ports, so the performance of the MFE lens and scattering parameters can be investigated more precisely. The PhC structure consists of the 2D triangular lattice of cylindrical *Si* rods surrounded by $SiO_2$ as host [55, 56]. In simulations, the refractive indices of *Si* and $SiO_2$ are considered as *3.45* and *1.45*, respectively. The lattice constant is $a = 465nm$ and rods have a radius of $r = 0.2a$ [57]. This PhC structure has a bandgap in TM mode covering S, C, L, and U optical communication bands. Although this paper focuses on designing of multilayer MFE lens for the TM mode, it is also possible to design the MFE lens for the TE mode with the method described in the previous section. In our design, to avoid reflection, the MFE lens's refractive index changes radially from *2.9* at the center to *1.45* at the edge of the lens to match the refractive index of the host material. The multilayer MFE lens is designed with the method described in the previous section. The approximation method from 3D model to 2D model is based on simple area averaging of Si and $SiO_2$ layers. This approximate method does not consider the evanescent mode in claddings, however, it has been validated in [34, 39, 58, 59].

### 3.1 Concentric ring-based MFE lens

The ring-based lens with the radius of $2.56\mu m$ is divided into *10* cylindrical layers. The average refractive index of each layer is mapped to the width of the inclusion layer determined by Eq. (5). The width of inclusion layers decreases towards the edge of the lens indicating the decrease of the effective refractive index of the MFE lens.

Fig. 2 illustrates the computational domain of the simulation. To limit the number of simulations, the symmetrical computational domain is employed. One of the important factors considered in the construction of the computational domain is its size. Therefore, different domain truncation methods such as the perfectly matched layer (PML) or scattering boundary condition (SBC) are used to truncate the computational domain. In order to reduce the spurious reflection, the PhC waveguides are terminated by PhC-based PML. The PML domain surrounded by PhC structure is called PhC-based PML [60]. PhC-based PML is more effective compared to the conventional PML because the wave in the PhC-based PML region experiences the same refractive index contrast as in the PhC region [60] and moreover, the shape of the wave is preserved reducing the reflection from the ports [61]. Truncating the waveguide ends to SBC instead of PML, distorts the scattering parameters calculation with substantial reflection from the ports.

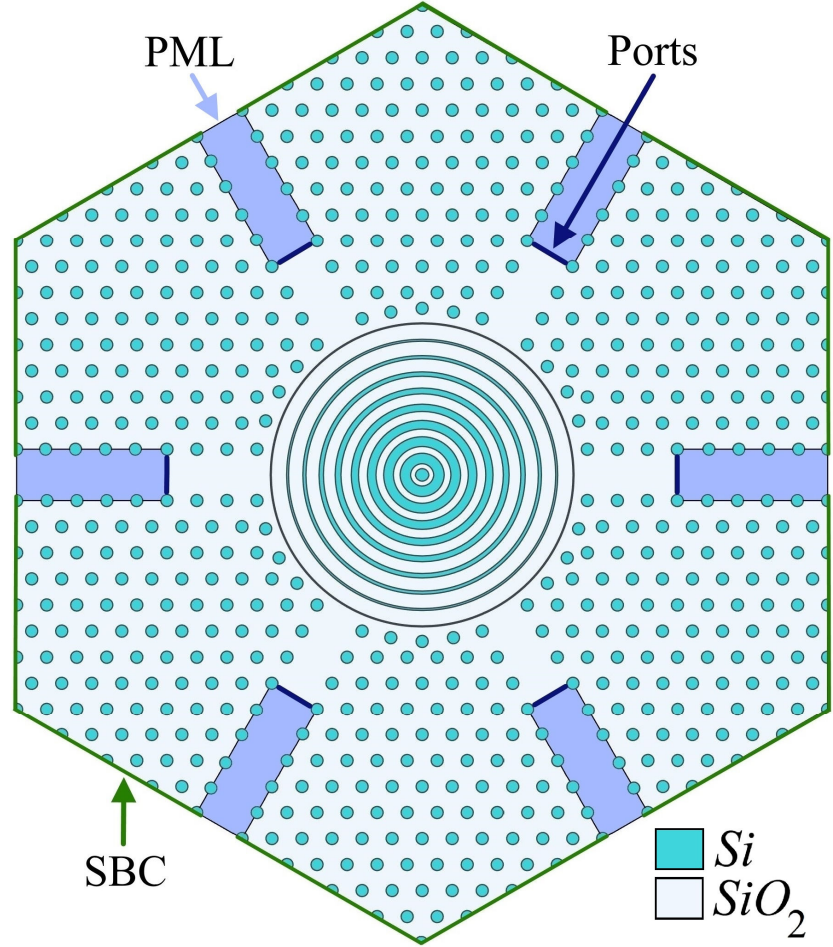

Fig. 2. Computational Domain. PhC-based PML and SBC are used to truncate the computational domain. All the rods and inclusion layers are Si and the host is SiO$_2$.

The transverse cross-section of the designed ring-based MFE lens is shown in Fig. 3(a). The filling factor of each inclusion layer is displayed numerically and graphically in the upper part of the Fig. 3(a). And the lower part of the Fig. 3(a) shows that the lens is divided into ten concentric cylindrical layers. The inclusion layers of Si are located in the center of cylindrical layers. Upper and lower claddings are SiO$_2$. Fig. 3(b) illustrates the propagation of TM mode signal at *1550nm* passing through the ring-based MFE lens where the return loss of *-14.2dB*, insertion loss of *0.05dB*, and crosstalk levels below *-20.5dB* are achieved. The waveguides are formed by removing one row of rods, so they only support TM$_0$. The scattering parameters of the ring-based 10-layer intersection are shown in Fig. 4(a). The eight transmission bands with a transmission of higher than -3dB and linear phase response are highlighted. The number of layers and the space between the last layer and the waveguides are optimized to cover the entire C-band. In the C-band, the return loss of below *-11.2dB*, the average insertion loss of *0.2dB*, and crosstalk levels below *-17.4dB* are achieved. Furthermore, the intersection covers about *80%* of the U-band. Four of the MFE lens's modes are also presented in Fig. 4(b). The mode B is the prevalent mode of the lens in the C-band with lower insertion loss and crosstalk levels. For the mode D, although the transmission is higher than *-3dB* the phase response is nonlinear and is not considered in the transmission band. The analytical study of the MFE lens's modes has been addressed in [62, 63]. The effect of device geometry errors was simulated to evaluate the tolerance of the concentric ring-based MFE lens to fabrication errors. The proposed structure can tolerate the variation in width of the inclusion layers of up to ±5$nm$ with the maximum excess insertion loss of 1.7$dB$ in the C-band.

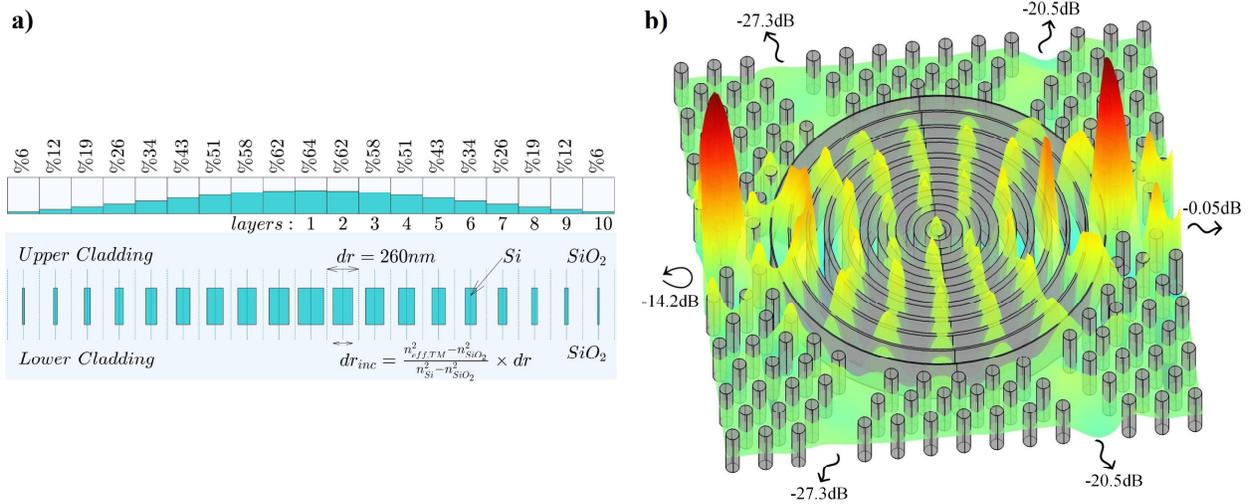

Fig. 3. Concentric ring-based MFE lens. a) The transverse cross-section of the designed ring-based MFE lens. The filling factor of each inclusion layer is displayed numerically and graphically in the upper part of this figure. And the lower part of this figure shows the inclusion layers of Si surrounded by SiO$_2$. The upper and lower cladding are SiO$_2$. b) Electric field distribution of 3×3 intersection based on concentric cylindrical multilayer MFE lens at 1550$nm$.

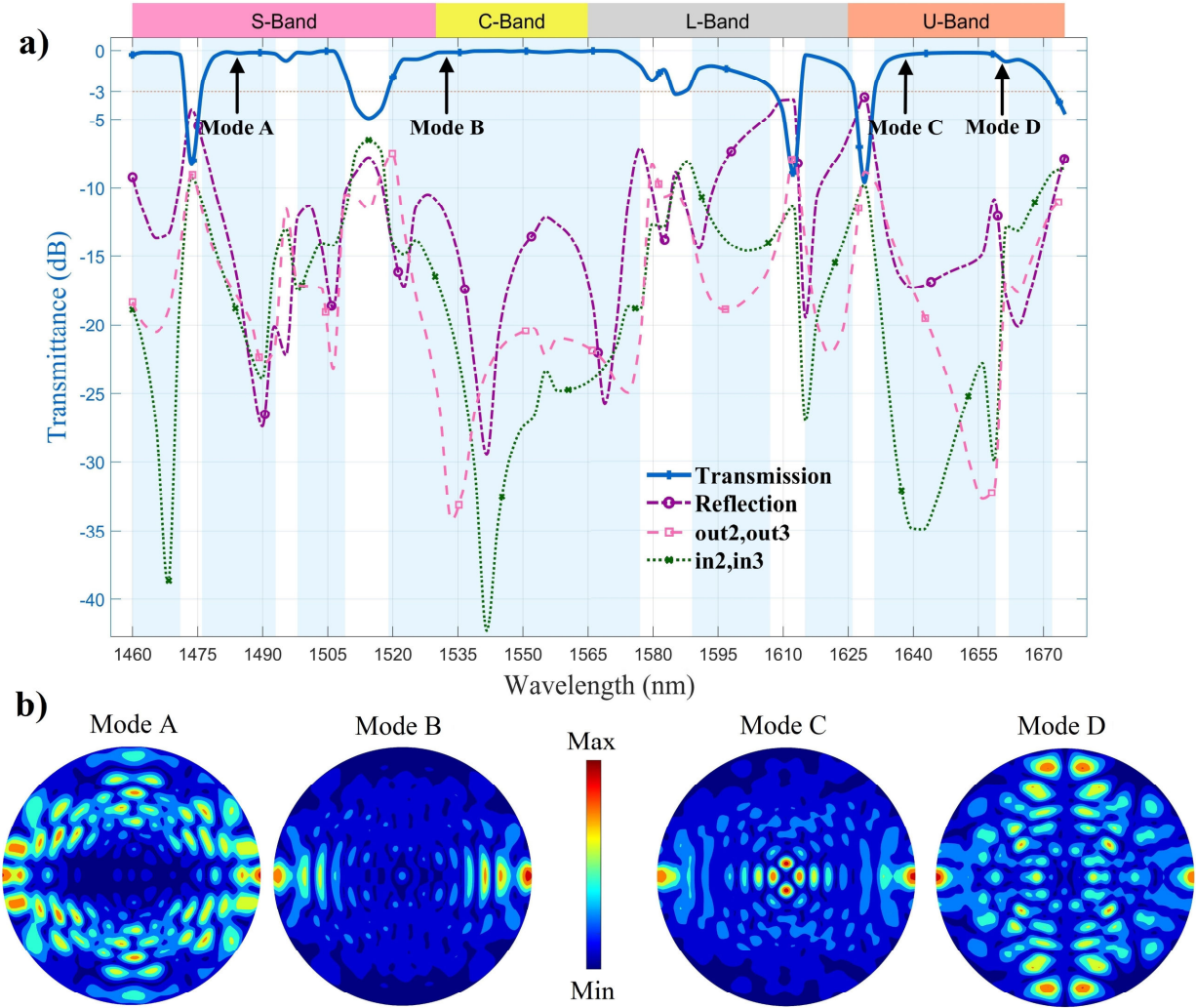

Fig. 4. Scattering parameters and modes. a) Scattering parameters of the designed ring-based multilayer intersection. b) Electric field distribution of selected modes of the MFE lens.

### 3.2 Radial-based MFE lens

An alternative approach in the implementation of the refractive index profile of the MFE lens is radial multilayers [50, 64]. The procedure described in the previous section is used for calculating the radial layers' geometry of the MFE lens with the radius of $2.56 \mu m$ illustrated in Fig.5 (a). The designed gourd-shaped radial rods of Si and ten annular layers are shown in this figure. The radial rods of Si are surrounded by $SiO_2$. The angle between consecutive radial rods is 7.5°. The first layer is designed similar to the ring-based design procedure resulting in a gap between the first layer and the radial rods. In order to satisfy the long-wavelength limit condition, the number of radial rods is increased towards the edge of the lens. As described in previous section, each layer is divided into a number of equal annular sectors. Moving from the center towards the edge of the lens, the number of annular sectors in each layer is doubled to double the number of rods in that layer.

However, to ease the fabrication limitations, after four layers the number of sectors is kept constant. Increasing the number of radial rods results in thinner rods and reduction of gourd shape sections' curvature, hence the fabrication constraints dictate the number of radial rods. The radial rod consists of gourd-shaped sections and one long graduated tip. The rods added in the second layer have triple gourd-shaped sections. However, the rods added in the third and fourth layers have double and single gourd-shaped sections. The rods added in the fifth layer only have a long graduated tip. The electric field distribution of the radial-based intersection is shown in the Fig. 5(b) at *1487nm*.

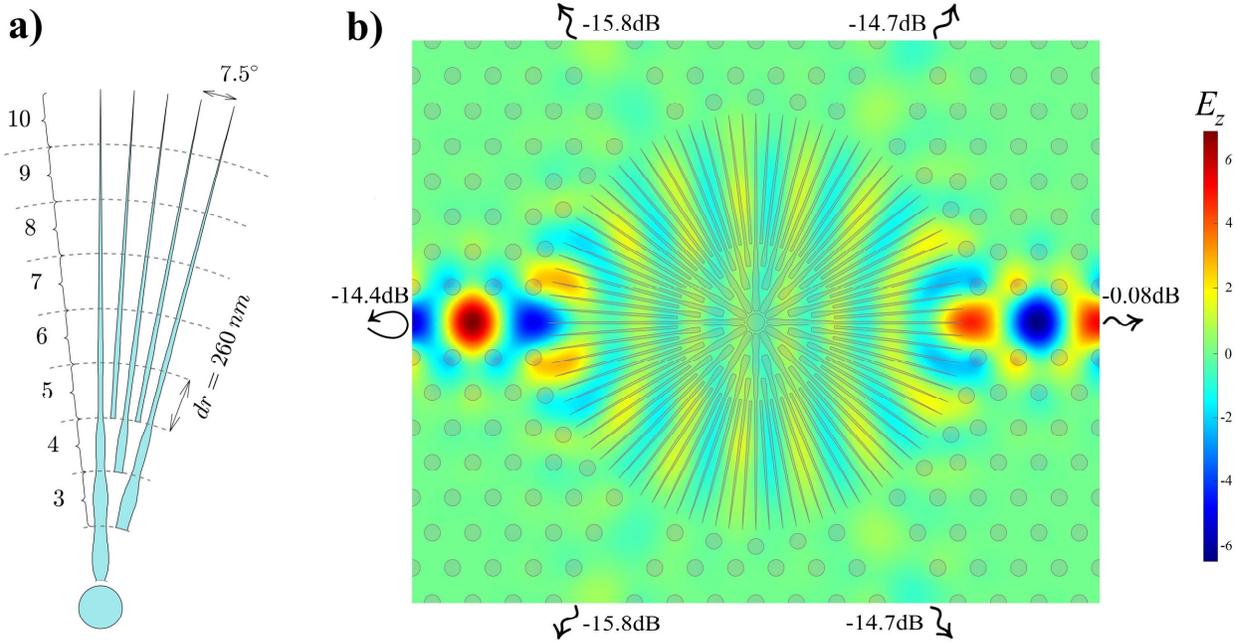

Fig. 5. Radial-based MFE lens. a) Geometry of a radial rods of Si and annular layers. The radial rods of Si are surrounded by $SiO_2$. b) Electric field distribution of radial-based multilayer intersection at *1487nm* .

The scattering parameters of the radial-based intersection are illustrated in Fig. 6(a). The six transmission bands with transmission higher than *-3dB* and linear phase response are highlighted. In Fig. 6(b), the radial-based lens's modes are presented which are different from the ring-based lens. The number of gourd-shaped radial rods in each layer is optimized to cover the entire C-band. In the C-band, the return loss of below *-4.3dB*, the average insertion loss of *1.5dB*, and crosstalk levels below -10.8dB are achieved. Contrary to the ring-based design, it covers the entire U-band but with higher insertion loss. It also covers about *58%* of the L-band. To provide a better understanding of the light pulse propagation through the radial-based MFE lens, a movie is presented at [65]. This video is created by Lumerical™ FDTD simulation software. The central frequency of the input pulse is *185.173THz* and its bandwidth is *41.3641THz*. In Fig. 7, the transmission capability of the two presented designs are compared. Both of the ring-based and radial-based multilayer intersections successfully cover the entire C-band. Although the ring-based lens has a better response in the C-band and lower insertion loss, the radial-based lens covers almost the entire U-band and has wider transmission bands. The radial-based MFE lens can tolerate the random variation in width of the radial rods of up to ±5*nm* with the maximum excess insertion loss of 1.1*dB* in the C-band.

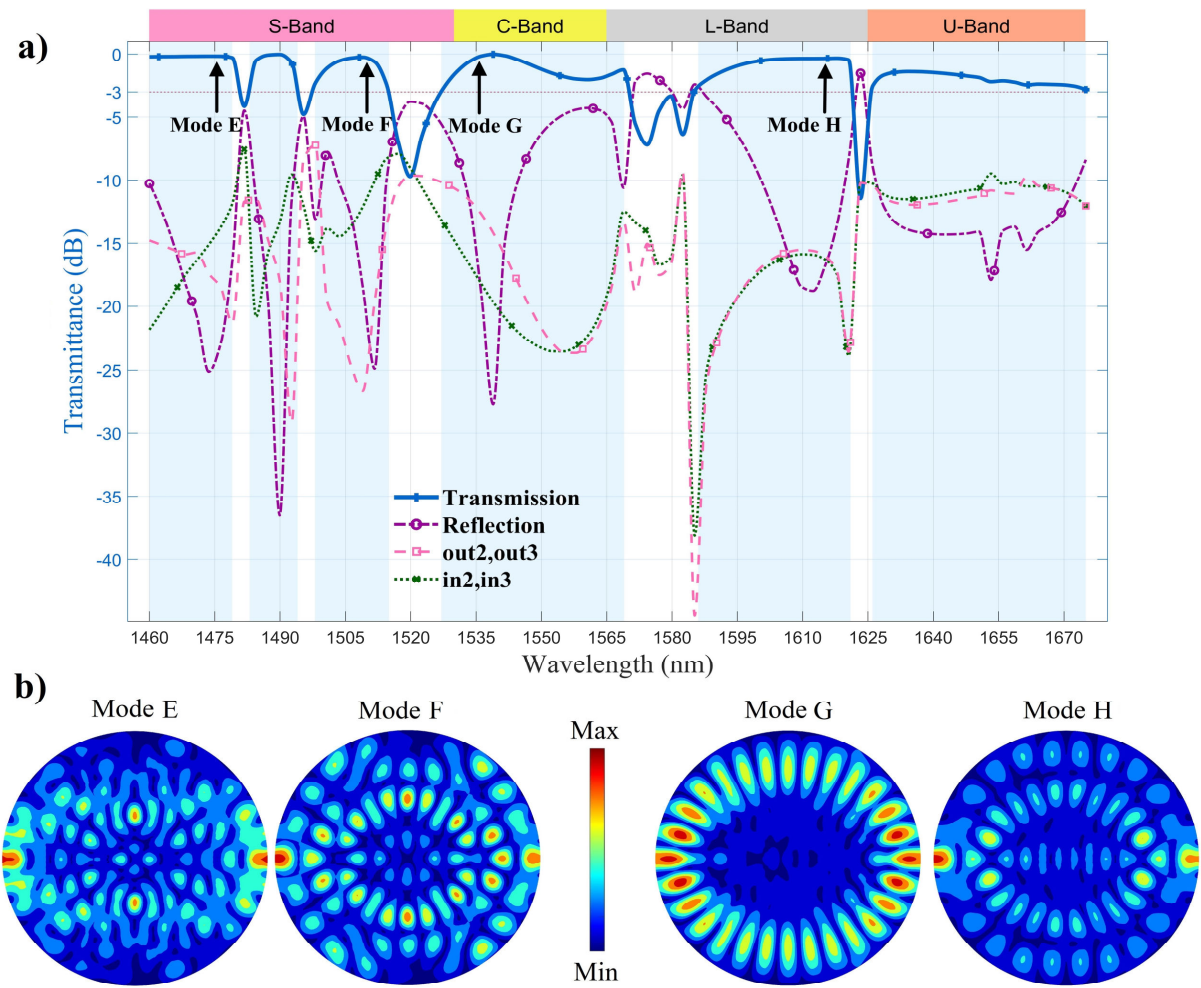

Fig. 6. Scattering parameters and modes. a) Scattering parameters of the designed radial-based multilayer intersection. b) Electric field distribution of selected modes of the MFE lens.

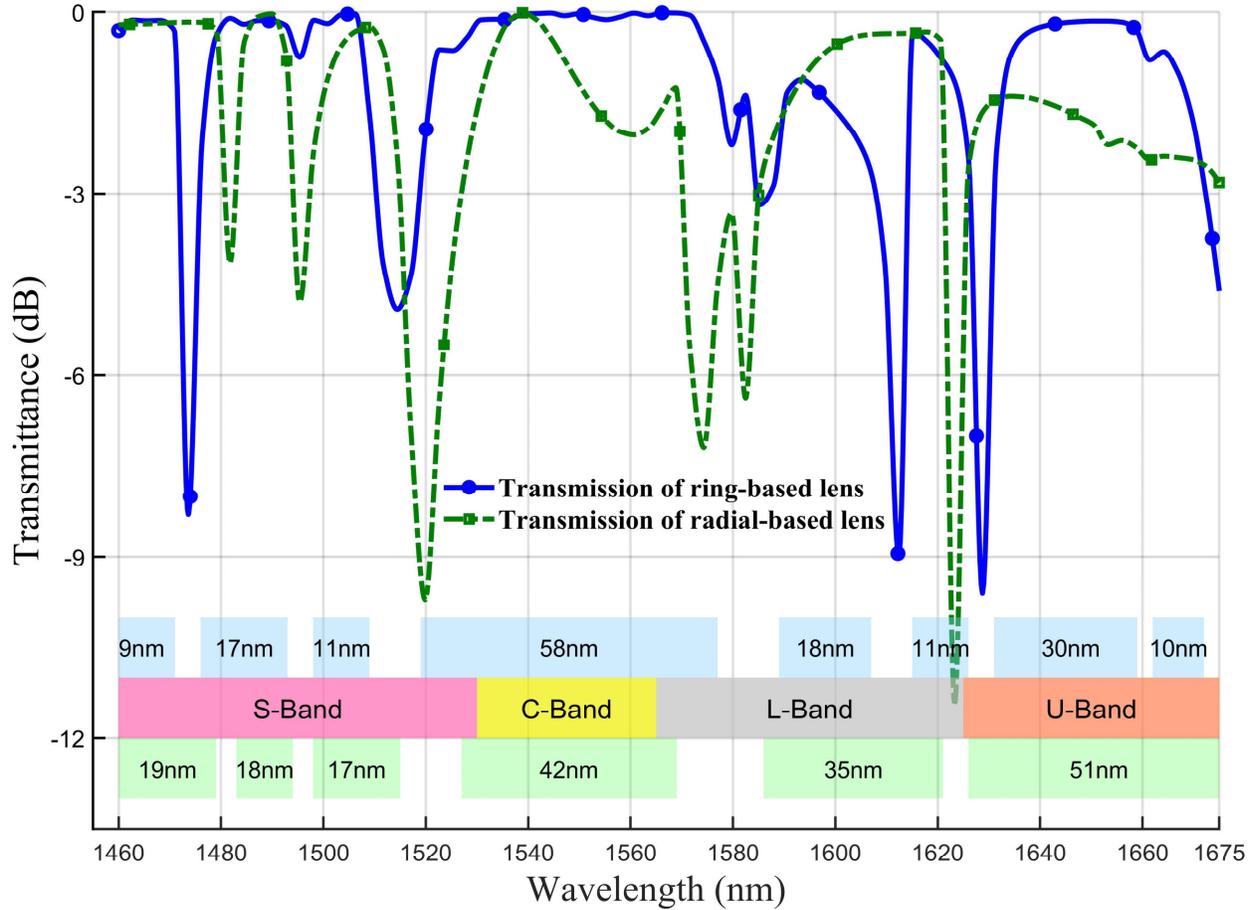

Fig. 7. Comparison of the ring-based and radial-based multilayer intersections' transmission. Blue and green horizontal bars specify the bands with the acceptable transmission for ring-based and radial-based intersections, respectively. Each transmission band's bandwidth is determined.

### 3.3 Multimode waveguide crossing

The MFE lens can also support the transmission of higher modes. In this paper, transmission of $TM_0$ and $TM_1$ through the ring-based MFE lens is also investigated. However, the simulations, not presented here, show that the MFE lens can support higher modes simply by increasing the waveguides' width and the lens's radius. The multimode photonic crystal waveguides (PhCW) are obtained by removing multiple neighboring rows of rods [18, 24, 66]. In our simulations, three consecutive rows are removed to form a multimode PhCW. The electric field distribution for $TM_0$ and $TM_1$ are illustrated in Fig. 8 at the wavelength of *1550nm*. The scattering parameters of the multimode waveguide intersection are illustrated in Fig. 9. As the width of the waveguide increases the optical confinement decreases, resulting in lower insertion loss and lower crosstalk levels [8]. The three transmission bands where both $TM_0$ and $TM_1$ have transmissions higher than *-3dB* and linear phase responses are highlighted. In the C-band, the return loss of below *-14.8dB*, the average insertion loss of *0.17dB*, and crosstalk levels below *-24.2dB* are achieved.

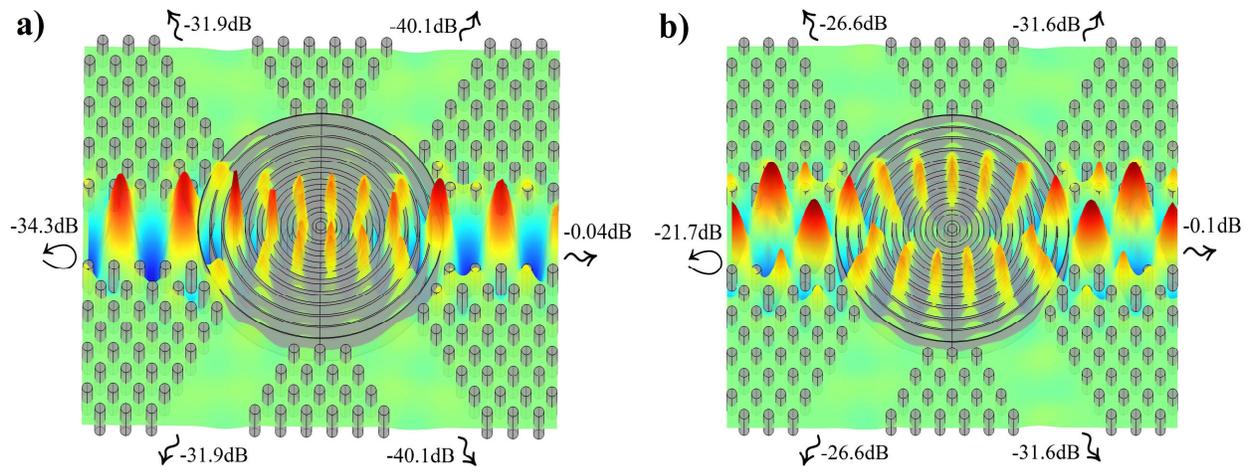

Fig. 8. Multimode ring-based multilayer intersection. The electric field distribution is illustrated at the wavelength of 1550*nm* for a) $TM_0$ and b) $TM_1$ modes.

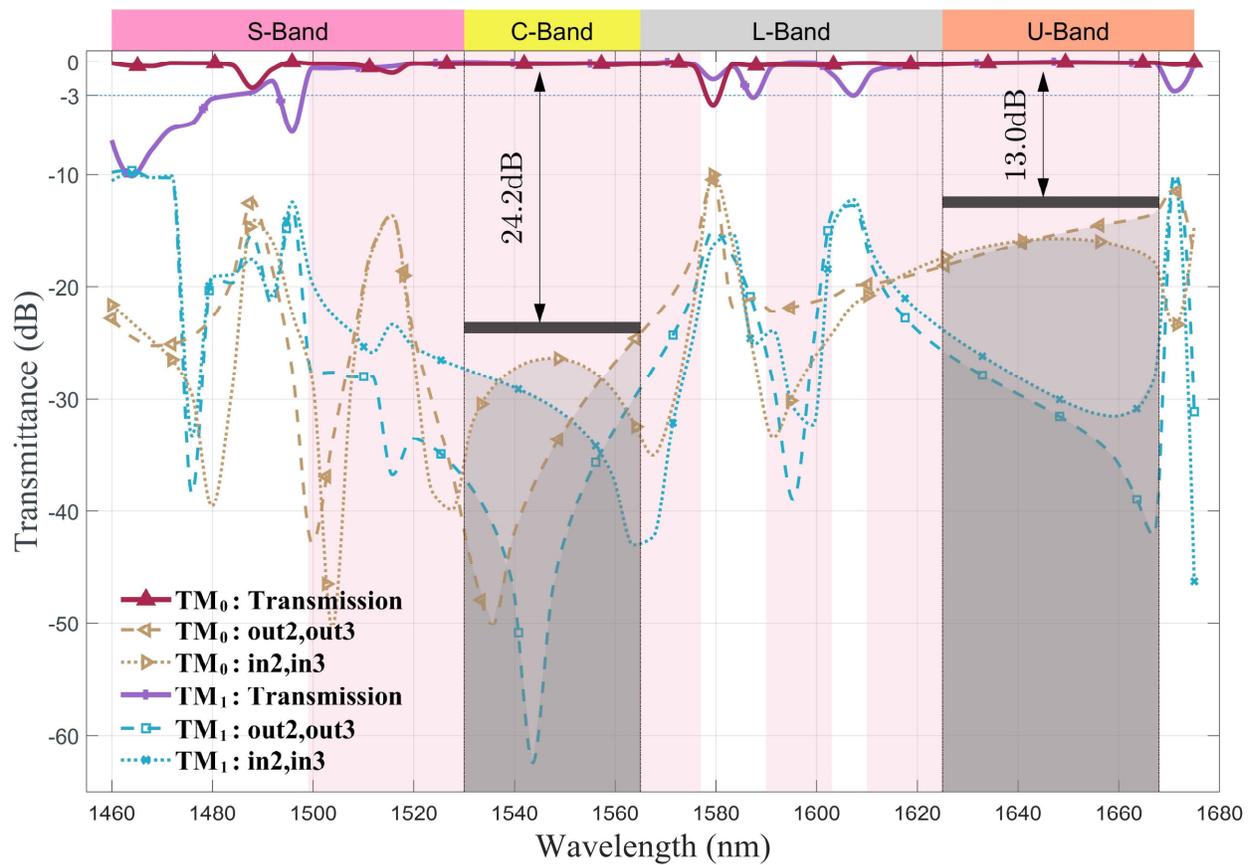

Fig. 9. Comparison of the multimode intersections' transmission. The transmission bands covering both $TM_0$ and $TM_1$ modes are highlighted.

The characteristics of the waveguide crossings of the references discussed in introduction section are summarized in Table 1. The crossing mechanism, insertion loss, central wavelength, bandwidth, crosstalk, and footprint are compared in this table. It has also been specified whether the intersection of multiple waveguides and intersection of multimode waveguides are reported in the literatures mentioned in the introduction section. Some of the specifications are not explicitly given in the references, therefore they are extracted from the figures. Some references used normalized frequency based on lattice constant. In order to provide tangible specifications, when the lattice constant was not given, it was calculated in a way that the center of the crossing's passband lies in the center of the photonic band gap. Only the crossings based on the MFE lens support the crossing of multiple multimode waveguides at an intersection. Compared to the other works, the footprint of the presented multimode waveguide crossing is smallest.

Tabel 1. Characteristics of the waveguide crossings

| reference | Year | Crossing Mechanism | Polarization | Waveguide Type | Insertion Loss (dB) | $\lambda_{center}$ (nm) | Bandwidth (nm) | Cross-talk(dB) | Footprint ($\mu m^2$) | intersection of multiple waveguides reported | intersection of multimode waveguides reported |
|---|---|---|---|---|---|---|---|---|---|---|---|
| [1] | 1998 | Cavity | TM | PhC | - | 1550 | 17 | -34 | 1.2×1.2 | No | No |
| [2] | 2002 | Cavity | TE | PhC | 6 | 1306 | 4 | -18 | 5.6×5.6 | No | No |
| [3] | 2007 | Cavity | TE | PhC | 0.91 | 1300 | 30 | -30 | 3.6×3.6 | No | No |
| [4] | 2006 | topology optimization | TE | PhC | 2 | 1275 | 50 | -22 | 1.8×2.2 | No | No |
| [5] | 2005 | Cavity | TE | PhC | 0.04 | 1550 | 33 | -40 | 2.7×2.7 | No | No |
| [6] | 2012 | Cavity | TM | PhC | 0.09 | 1540 | 4.66 | -18 | 3.65×3.65 | No | No |
| [7] | 2006 | Multimode-interference | TE | Si waveguide | 0.4 | 1550 | 100 | -33 | 13×13 | No | No |
| [8] | 2010 | Multimode-interference | TE | Si waveguide | 0.21 | 1550 | 100 | -38 | 5.5×5.5 | No | No |
| [9] | 2007 | Mode expander | TE | Si waveguide | 0.16 | 1550 | 40 | -40 | 6×6 | No | No |
| [10] | 2007 | Wavefront matching | TE/TM | Silica waveguide | 0.35 | 1550 | - | -35 | 120×230 | No | No |
| [11] | 2010 | impedance matched metamaterials | TM | Si waveguide | 0.04 | - | - | -40 | - | No | No |
| [12] | 2017 | Inverse-designed | TE | Silica waveguide | 0.75 | 1550 | 80 | -22.5 | 5.33×5.33 | Yes | No |
| [13] | 2018 | MFE lens | TM | PhC | 0.6 , 1.3 | 1561 , 1643 | 64 , 63 | -16 , -18 | 7.6×7.6 | Yes | No |
| [31] | 2018 | Multimode-interference | TE | Si waveguide | 0.6 | 1560 | 60 | -24 | 4.8×4.8 | No | Yes |
| [32] | 2016 | Multimode-interference | TM | Si waveguide | 1.5 | 1560 | 80 | -18 | 29×29 | No | Yes |
| [33] | 2018 | Asymmetric Y-junction | TE | Si waveguide | 0.9 | 1560 | 80 | -24 | 34×34 | No | Yes |
| [34] | 2018 | MFE lens | TM | Si waveguide | 0.3 | 1560 | 80 | -20 | 14×14 | Yes | Yes |
| radial-based MFE lens | this work | MFE lens | TM | PhC | 1.1 , 2.1 | 1552 , 1649 | 42 , 51 | -10 , -9 | 5.1×5.1 | No | Yes |
| concentric ring-based MFE lens | this work | MFE lens | TM | PhC | 0.2 , 0.15 | 1538 , 1638 | 78 , 57 | -14 , -13 | 4.65×4.65 | Yes | Yes |

## 4. Conclusion

The MFE lens, due to its imaging properties, is utilized as the intersection of multiple waveguides supporting any number of allowed modes. We believe that for the first time the MFE lens is implemented as ring-based and radial-based multilayer metamaterial comprising periodic Si and SiO$_2$ layers in this work. The simulations show that the all-dielectric metamaterial-based MFE lens can be used to implement broadband single-mode and multimode waveguide crossings with low insertion loss and low crosstalk levels. For the single-mode waveguide crossing, crosstalk levels below *-17.4dB* and the average insertion loss of *0.2dB*, are achieved in the C-band. For the multimode waveguide crossing based on the MFE lens with $4.65 \times 4.65 \mu m^2$ footprint, crosstalk levels lower than *-24.2dB* and the average insertion loss of *0.17dB* are obtained for both TM$_0$ and TM$_1$ modes in the C-band.